\documentclass[twocolumn,preprintnumbers,amsmath,amssymb,nofootinbib]{revtex4-1}
\usepackage[dvipdfmx]{graphicx}
\usepackage{here}
\usepackage{comment,braket}
\usepackage{epsf}
\usepackage{amsmath}
\usepackage{graphics}
\usepackage{amsfonts}
\usepackage{amssymb}
\usepackage{latexsym}
\usepackage{color}
\usepackage{natbib}
\usepackage{hyperref}
\usepackage[svgnames]{xcolor}
\definecolor{phthaloblue}{rgb}{0.0, 0.06, 0.54}
\hypersetup{
    colorlinks=true,
    linkcolor=black,
    citecolor=red,
    filecolor=blue,
    urlcolor=black,
    }
\usepackage{feynmp}
\begin{document}

\title{Small-mass naked singularities censored by the Higgs field}

\author{Naritaka Oshita}
\affiliation{
Perimeter Institute for Theoretical Physics, Waterloo, Ontario N2L 2Y5, Canada
}

\begin{abstract}
We investigate a vacuum decay around an over-spinning naked singularity by using the Israel junction condition. We found that if the Higgs field develops the second minimum at higher energy scale, a spinning small-mass naked singularity could cause the vacuum decay around it within the cosmic age. An event horizon may form around the singularity due to the angular momentum transport from the singularity to a vacuum bubble wall. The newly formed event horizon leads to the increase of Bekenstein-Hawking entropy, which contributes to the enhancement of the vacuum decay rate. We conclude that small-mass naked singularities may be hidden by the event horizon within the cosmological time.
\end{abstract}

\maketitle

\section{Introduction}
The weak cosmic censorship conjecture \cite{1974AnPhy82548W} states that any naked singularities, except for the Big Bang singularity, cannot exist in the Universe and those singularities should be hidden by event horizons. The censorship mechanism to prevent the naked singularities to exist in the Universe has been investigated mainly in a classical manner so far. For instance, it was pointed out that a nearly-extremal black hole (BH) could be a naked singularity by capturing a particle which carries a small amount of angular momentum \cite{Jacobson:2009kt}, provided that relevant backreaction effects are negligible \cite{PhysRevD.91.104024,Colleoni:2015ena}. On the other hand, assuming a priori that a spinning naked singularity exists in the Universe, the life-time of a spinning naked singularity surrounded by stellar medium was estimated by taking into account matter accretion to the singularity \cite{1975AA4565D,deFelice:2007fs}. The estimated life-time of the naked singularity is
$T \sim 10^4 \left({M_{\odot}/M_+} \right) \left(10^{-13} \text{g cm}^{-3}/\rho\right) \ \text{years}$,
where $M_+$ is the mass of the naked singularity and $\rho$ is the density of accreting particles. This life-time is inversely proportional to $M_+$, and so
if we consider a small-mass naked singularity, the life-time may be longer than the cosmic age $\sim 1.38 \times 10^{10}$ years. Therefore the accretion to naked singularities would not provide the censorship mechanism and the cosmic censorship conjecture may be drastically violated especially for the small-mass naked singularity.

In this {\it Letter} we propose a novel scenario to explain the censorship mechanism for such a small-mass naked singularity from the point of view of a semi-classical picture. It has been well known that the extrapolation of the standard model up to very high energy scale would lead to the possible Higgs metastability \cite{
Sher:1988mj,Arnold:1989cb,Altarelli:1994rb,Espinosa:1995se,Casas:1996aq,Hambye:1996wb,Isidori:2001bm,Espinosa:2007qp,Ellis:2009tp,Bezrukov:2012sa,Bednyakov:2015sca,EliasMiro:2011aa,Degrassi:2012ry,Buttazzo:2013uya,Branchina:2013jra}, and the vacuum decay due to this metastability may be significantly promoted by microscopic BHs \cite{PhysRevD.35.1161,Gregory:2013hja,Burda:2015yfa,Burda:2015isa,Burda:2016mou,Gregory:2016xix,Oshita:2019jan} or horizon-less compact objects \cite{Steinhardt:1981ec,Oshita:2018ptr,Koga:2019mee}. Since impurities usually catalyze phase transitions, there is no wonder that the Higgs metastability may be catastrophic around such dense objects.
\begin{figure}[h]
  \begin{centering}
    \includegraphics[width=0.45\textwidth]{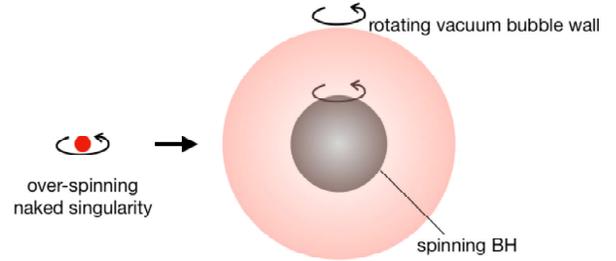}
  \end{centering}
\caption{A schematic picture showing the formation of event horizon around the singularity due to the angular momentum transport to the vacuum bubble from the naked singularity.
}
\label{cartoon}
\end{figure}

We will consider vacuum decay around a spinning naked singularity using the thin-wall approximation, and will show that the vacuum decay involves the formation of event horizon around the singularity due to the angular momentum transport to a vacuum bubble wall (Fig. \ref{cartoon}). This leads to the increase of Bekenstein-Hawking entropy \cite{Bekenstein:1973ur,Bekenstein:1974ax,Bekenstein:1975tw,Hawking:1976de} in the system, which contributes to the enhancement of the decay rate. In the following we use the natural unit $\hbar = c= k_B =1$.

\section{Dynamics of vacuum bubble}
Here we calculate a vacuum decay rate around a highly spinning naked singularity of mass $M_+$. Especially, we here consider the first order phase transition by which the angular momentum of a seed naked singularity is transferred to a vacuum bubble wall, {and the event horizon forms around the singularity. To this end, we will calculate the rate of a transition from a spinning naked singularity (initial state) to a thin-wall vacuum bubble with a Kerr anti-de Sitter (Kerr-AdS) interior (final state) by using the Euclidean path integral. Then we will compare it with the cosmic age.}
Note that here we implicitly assume that the difference in the vacuum energy density between the true and false vacuum state is much smaller than the height of the barrier of the Higgs potential\footnote{For instance, to model such a Higgs potential, sextic term of Higgs field was introduced in \cite{Burda:2015yfa}.}. This assumption is necessary to guarantee that the resulting vacuum bubble has its ``thin" wall. Since here we consider a thin-wall bubble as a final state, the resulting spacetime may be described by the Israel junction condition between the Kerr-AdS and Kerr spacetime. Using the Boyer-Lindquist (BL) coordinates, a metric which covers the Kerr-AdS spacetime has the form of
\begin{widetext}
\begin{equation}
ds^2 = - \frac{\Delta_r}{\Sigma} \left( dt - \frac{a \sin^2{\theta}}{\Xi} d\phi \right)^2 + \frac{\Sigma}{\Delta_r} dr^2 + \frac{\Sigma}{\Delta_{\theta}} d \theta^2 + \frac{\Delta_{\theta} \sin^2{\theta}}{\Sigma} \left(  a dt - \frac{r^2 + a^2}{\Xi} d \phi \right)^2,
\label{BLmetric}
\end{equation}
\end{widetext}
where $\Sigma \equiv r^2 + a^2 \cos^2 {\theta}$, $\Xi \equiv 1 + H^2 r^2$, $\Delta_r \equiv (r^2+a^2) (1+ H^2 r^2) - 2GM r$ and $\Delta_{\theta} \equiv 1+H^2 a^2 \cos^2{\theta}$, $M$ is the singularity mass, $a$ is the spin parameter, and $H$ is the Hubble parameter. The spacetime inside and outside the wall can be obtained by setting $a = a_+ > GM_+$, $H = H_+ = 0$ and $M = M_+$ in (\ref{BLmetric}), and the interior spacetime is also given by (\ref{BLmetric}) with $a = a_- \geq 0$, $M = M_-$, and $H = H_- > 0$, where $M_-$ is the remnant mass and $H_-$ is the Hubble parameter at the true vacuum of Higgs field. Suffixes of $+$ and $-$ represent the exterior and interior quantities.  Let us assume that the radius of a nucleated vacuum bubble is much larger than $2GM_{\pm}$ and much smaller than $1/H_-$. In this case the metric near the wall can be approximated by
\begin{align}
\begin{split}
ds^2 \simeq& - f_{\pm}(r_{\pm}) dt_{\pm}^2 + \frac{dr_{\pm}^2}{f_{\pm}(r_{\pm})} + r_{\pm}^2 d\Omega^2_2\\
&- \frac{2 r_{\text{s} \pm} a_{\pm} \sin^2{\theta}}{r_{\pm}} dt_{\pm} d\phi,
\end{split}
\label{slowBL}
\end{align}
where $f_{\pm}(r) \equiv 1- r_{\text{s} \pm}/r + H_{\pm}^2 r^2$ and $r_{\text{s}}{}_{\pm} \equiv 2GM_{\pm}$. Here we ignore the terms of the order of ${\mathcal O} (a_{\pm}^2/r_{\pm}^2)$ in (\ref{BLmetric}). One can eliminate the off-diagonal term in (\ref{slowBL}) by going to the co-rotating (static) frame with the coordinate transformation of $d\phi \to d\psi + (r_{\text{s} \pm} a_{\pm}/r_{\pm}^3) d t$ \cite{Poisson:2009pwt}, and the static metric is given by
\begin{align}
\begin{split}
ds^2 &\simeq g_{\pm \mu \nu} dx_{\pm}^{\mu} dx_{\pm}^{\nu}\\
&= - \left(f_{\pm} + \frac{r_{\text{s}\pm}^2 a_{\pm}^2 \sin^2{\theta}}{r_{\pm}^4} \right)dt_{\pm}^2 + \frac{dr_{\pm}^2}{f_{\pm}} + r_{\pm}^2 d\Omega_2^2.
\end{split}
\label{static}
\end{align}
{When we consider a vacuum bubble whose interior and exterior surfaces, denoted by $\Sigma_{\cal W +}$ and $\Sigma_{\cal W -}$, respectively, are given by
\begin{equation}
\Sigma_{{\cal W} \pm} = \left\{ (t_{\pm}, r_{\pm}, \theta_{\pm}, \phi_{\pm}) | r_{\pm}-R (\tau (t_{\pm}))=0 \right\},
\label{surface_shell}
\end{equation}
the induced metric on $\Sigma_{\cal W \pm}$ are
\begin{align}
\begin{split}
ds^2 &= \left\{ - \left( f_{\pm} + \frac{r_{\rm s \pm}^2 a_{\pm}^2 \sin^2 \theta}{R^4} \right) \dot{t}_{\pm}^2 + \frac{\dot{R}^2}{f_{\pm}} \right\} d\lambda^2\\
&+ R^2 d\theta^2 + R^2 \sin^2 \theta d \psi^2,
\end{split}\\
& = - d\lambda^2 + R^2 d\theta^2 + R^2 \sin^2 \theta d\psi^2,
\label{induced_met}
\end{align}
where we used $-1 = g_{\pm 00} dt_{\pm}^2 + g_{\pm 11} dr_{\pm}^2$. The first Israel junction condition requires the continuity of the interior and exterior induced metrics, which is {approximately} satisfied as can be seen from (\ref{induced_met}). {This approximation is valid only when $a_{\pm}^2 / r_{\pm}^2 \ll 1$ and so in this {\textit Letter} we only consider a case where a vacuum bubble is much larger than the Schwarzschild radius $r_{s \pm}$. For a more rigorous error estimation in the second Israel junction condition due to this approximation, see the Appendix A.}
Let us next investigate the dynamics of the nucleated bubble by using the second Israel junction condition} between the exterior and interior geometry. In general the radius of the wall is dynamical and its radius $R$ is a function of the proper time on the wall $R = R(\lambda)$. 
Using the ($t, \psi$)-component of the Israel junction condition, one can confirm that the total angular momentum of the system is conserved before and after the phase transition.  Since $\dot{R} = 0$ at the moment when the wall is nucleated, the $(t, \psi)$-component of the extrinsic curvature is given by \cite{Poisson:2009pwt}
\begin{equation}
K^{(\pm)}_{t \psi} \simeq \frac{3 r_{\text{s} \pm}}{2 R^2} a_{\pm} \sin^2{\theta},
\end{equation}
where we used $r_{\text{s} \pm} \ll R \ll 1/H_{-}$ {and $\psi$ denotes the azimuthal angle in the co-rotating frame}. The angular component of energy flux of the wall $S_{t \psi}$ is 
\begin{equation}
S_{t \psi} = \sigma R^2 \sin^2{\theta} \dot{\phi},
\end{equation}
where a dot denotes the derivative with respect to the proper time of the wall $\lambda$, and $\sigma$ is the energy density of the wall.
The Israel junction condition, $K_{t \psi}^{(+)}-K_{t \psi}^{(-)} = 8 \pi G S_{t \psi}$, leads to
\begin{equation}
\dot{\phi} = \frac{3 (r_{\text{s}+} a_+ -r_{\text{s}-} a_-)}{16 \pi G \sigma R^4}.
\end{equation}
The angular momentum of the spinning wall is given by
\begin{equation}
L = \int^{\pi}_{0} d\theta 2 \pi \sin^3{\theta} \sigma R^4 \dot{\phi} = M_+ a_+ - M_- a_-,
\end{equation}
which is nothing but the conservation of the angular momentum.
\begin{figure}[b]
    \includegraphics[width=0.48\textwidth]{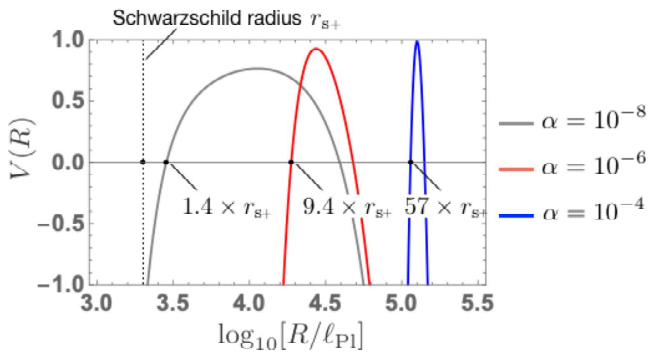}
\caption{The potentials governing the position of the bubble wall with $\alpha = 10^{-8}, 10^{-6}$, and $10^{-4}$. The mass of the seed naked singularity is fixed as $M_+ = 10^3 M_{\text{Pl}}$.
}
\label{potential}
\end{figure}

Using the $(\theta, \theta)$-component of the Israel junction condition, one can investigate the radial motion of the wall. The $(\theta, \theta)$-component of the extrinsic curvature is given by
\begin{equation}
K^{(\pm)}_{\theta \theta} = R \dot{t}_{\pm} f_{\pm} \sqrt{ 1 + \epsilon_{\pm}^4 \frac{\tilde{a}_{\pm}^2}{4 f_{\pm}} \sin^2{\theta}},
\label{etrinsicO}
\end{equation}
where $\epsilon_{\pm} \equiv r_{\text{s} \pm}/R$, and $\tilde{a}_{\pm} \equiv a_{\pm}/GM_{\pm}$. The rotation effect, whose term is of the order of $\epsilon_{\pm}^4$, can be ignored when $r_{\text{s} \pm} \ll R$. Therefore, as long as we consider vacuum decays leading to the nucleation of large vacuum bubbles such that $\epsilon_{\pm}^4 \ll 1$, we can ignore the rotation effect in (\ref{etrinsicO}) even for $\tilde{a}_{+} \gtrsim 1$. Then, the $(\theta, \theta)$-component of the Israel junction condition reduces to
\begin{equation}
\frac{\dot{t}_+ f_+ (R) - \dot{t}_- f_- (R)}{R} = - 4 \pi G \sigma,
\label{junc_thth}
\end{equation}
and eliminating $\dot{t}_{\pm}$ from (\ref{junc_thth}) one obtains
\begin{align}
&\dot{R}^2 + V (R) = 0,\label{dynamics_wall}\\
&V (R) \equiv 1- \frac{2GM_+}{R} - \left( \frac{2 G \Delta M/R^3 + H_-^2 - \Sigma^2}{2 \Sigma} \right)^2 R^2,
\end{align}
where $\Delta M \equiv M_+ - M_-$ and $\Sigma \equiv 4 \pi G \sigma$. This describes the spherical symmetric dynamics of the bubble wall. The radius of the nucleated bubble, $R_0$, satisfies $V(R_0)=0$ and we have two solutions for $R_0$ (see Fig. \ref{potential}). The smaller bubble ($R=R_{0}^{\text{min}}$) corresponds to the decaying mode and the larger one ($R=R_{0}^{\text{max}}$) is growing mode which is our interest here. Note that a condition
\begin{equation}
r_{\text{s} \pm}^2 \ll (R_0^{\text{min}})^2
\label{condition1}
\end{equation}
should be satisfied so that $r_{\text{s} \pm}^2 \ll R^2$ holds in the whole dynamics of the Euclidean bubble wall. Motivated by the Higgs vacuum decay, we take $H = 10^{-8} M_{\text{Pl}}$ and $H / \Sigma = 5000$ \cite{Burda:2015yfa} throughout the \textit{Letter}.  Let us confirm if the condition (\ref{condition1}) is satisfied in this set up. The mass of the singularity could change due to the gravitational transition \cite{Gregory:2013hja}, and so we will calculate the transition rate with various values of $\alpha \equiv - \Delta M/M_+$. One finds that a large value of $\alpha$ leads to a large-size vacuum bubble compared to the Schwarzschild radius  (Fig. \ref{potential}). Restricting to the naked singularity of $a_+ \gtrsim GM_+$, the parameter region of $\alpha \gtrsim 10^{-6}$ gives $\epsilon_+ \lesssim 10^{-4}$. In the following, we therefore calculate the decay rate only for the case of $a_+ \gtrsim GM$ and $\alpha \geq 10^{-6}$ due to the limitation of (\ref{condition1}).

\section{Higgs metastability around a spinning singularity}Performing the Wick rotation, $\lambda_E \equiv -i \lambda$ and $\tau_{\pm} \equiv - i t_{\pm}$, in the static metric (\ref{static}), one can calculate the on-shell Euclidean action, $I_{\text{E}}$, that is the exponent of the decay rate \cite{Gregory:2013hja}
\begin{align}
\begin{split}
I_{\text{E}}&\simeq \oint \frac{d \lambda_E}{4 G} \left[ (2R f_+-R^2 f'_+) \dot{\tau}_+ \right.\\
&~~~~~~\left. - (2R f_- - R^2 f'_-) \dot{\tau}_- \right] - \frac{\Delta A}{4G},
\end{split}
\label{euclidean}
\end{align}
where the integral part corresponds to the contribution from the Euclidean dynamics of the wall, and $\Delta A$ is the change of the horizon area before and after the phase transition. We are forced to assume that the naked singularity has no entropy to contribute to the Euclidean action since the naked singularity has no horizon by definition. Although the detail of the naked singularity is expected to be described by unknown physics such as quantum gravity, we believe that it is a reasonable assumption if the Planckian area is the smallest area that can accommodate at most one bit of information, and the naked singularity has its size comparable to or smaller than the Planckian size. 
\begin{figure}[b]
    \includegraphics[width=0.4\textwidth]{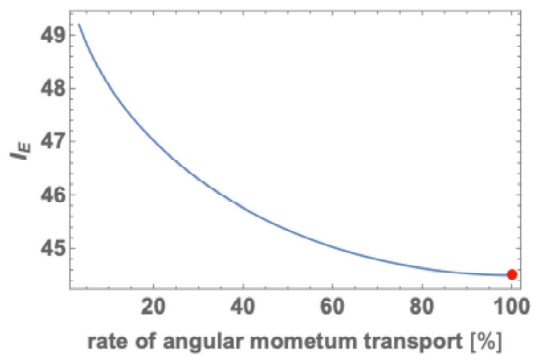}
\caption{The Euclidean action as a function of the rate of the angular momentum transport from the seed naked singularity to the vacuum bubble wall. The red point shows the minimum value of the action. We fixed $M_+ = M_{\text{Pl}}$, $\tilde{a}_+ = 1.001$, and $\alpha = 0.01 \%$.
}
\label{transport}
\end{figure}
\begin{figure*}[t]
    \includegraphics[width=1\textwidth]{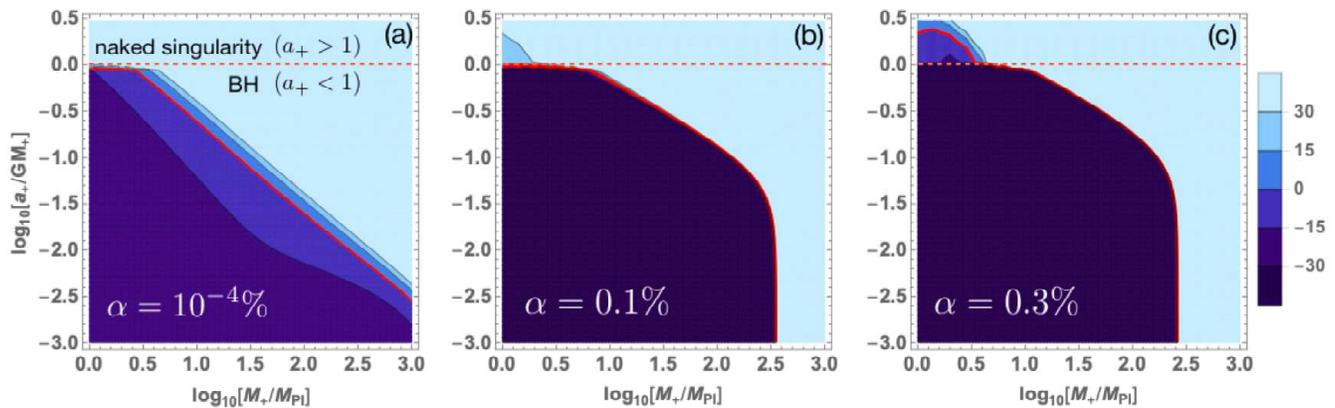}
\caption{Contour plots of $\log_{10}[t_c \times \Gamma]$ with (a) $\alpha = 10^{-4} \%$, (b) $\alpha = 0.1 \%$, and (c) $\alpha = 0.3 \%$. If $t_c \Gamma > 1$ (red lines show $t_c \Gamma = 1$), the vacuum decay may occur within the cosmological timescale.
}
\label{spin_rate}
\end{figure*}

However, the Planckian-size spinning singularity would have its spinning energy which could contribute to the Euclidean action since the on-shell Euclidean action can be regarded as free energy $E$ multiplied by a Euclidean time period $\tau_p$. Now we can roughly estimate the contribution of the singularity boundary to the action, given the angular momentum $M_+ a_+ \sim G M_+^2$, angular speed $\sim  \ell_{\text{Pl}} / a_+^2$ (c.f. \cite{Poisson:2009pwt}), and the Euclidean time period $\sim \ell_{\rm Pl}$ of the naked singularity\footnote{Here we assume that the naked singularity has its Planckian size.}. In this case, the contribution to the action is $\sim \tau_p E$:
\begin{equation}
\tau_p E \sim \ell_{\rm Pl} \times \frac{1}{2} M_+ a_+ \frac{\ell_{\rm Pl}}{a_+^2} = {\cal O} (1).
\end{equation}
Therefore, as long as the size of the naked singlarity is comparable with the Planckian length, the estimated contribution to the action is of the order of unity and we can safely ignore the contribution, provided that the semi-classical approximation is valid, i.e. $I_{\rm E} \gg 1$.

Moreover, the ignorance of the kinetic energy of the singularity gives an underestimate (conservative estimation) of the transition rate.
The Euclidean dynamics can be obtained by changing the sign of $\dot{R}^2$ in (\ref{dynamics_wall}), that is, the Euclidean bubble is the oscillatory dynamics under the effective potential of $U(R) \equiv - V(R)$. The second term tells us that the increase of the Bekenstein-Hawking entropy enhances the transition rate $\Gamma$ since it has the form of
\begin{equation}
\Gamma \sim \frac{1}{r_{\text{s}+}} \sqrt{\frac{I_{\text{E}}}{2 \pi}}e^{-I_{\text{E}}} \propto e^{\Delta A/4G},
\end{equation}
where we estimate the prefactor by taking a factor of $\sqrt{I_{\text{E}}/2\pi}$ for the zero mode associated with the time translation of $O(3)$ instanton, and we use the crossing timescale of over-spinning BH, $r_{\text{s}+}$, as a rough estimate of the determinant of fluctuations \cite{Gregory:2013hja}.
A vacuum decay, by which the angular momentum of the naked singularity is transferred to the bubble wall, would be catalyzed since the loss of angular momentum of singularity forms the event horizon and the Bekenstein-Hawking entropy increases. To show this, we calculate the Euclidean action (\ref{euclidean}) by changing the rate of the angular momentum transport (Fig. \ref{transport}). This shows that the phase transition by which all angular momentum of the singularity is transferred to the bubble wall is the most likely to occur, and so in the following we restrict ourselves to this case. Such a phase transition hides the naked singularity behind the newly formed event horizon, and so the Higgs vacuum decay around the spinning naked singularity may be regarded as a censorship mechanism.

We calculate the transition rate in the parameter range of $0.001 \leq a_{+} \leq 3$, $\alpha \geq 10^{-6}$, $M_{\text{Pl}} \leq M_+ \leq 10^3 M_{\text{Pl}}$ and $a_- =0$ (Fig. \ref{spin_rate}). We then compare it to the relevant timescale $t_c$, which is the cosmic age/BH evaporation time \footnote{Here we take the evaporation timescale of $t_c = 10^4 \times G^2 M_{+}^3$. The spin-dependence of the evaporation time \cite{Page:1976ki} can be negligible since the Euclidean action is more dominant in the decay rate.} for $a_+ > 1$ / $a_+ < 1$, respectively. It is found that a naked singularity catalyzes the phase transition within the cosmic age for $\alpha \lesssim 10^{-3}$ (see Fig. \ref{spin_rate}-(a,b) for the case of $\alpha = 10^{-6}$ and $\alpha = 10^{-3}$). When $\alpha \gtrsim 10^{-3}$, a Planckian naked singularity can be long-lived than the cosmic age (see Fig. \ref{spin_rate}-(c) for the case of $\alpha = 0.003$).
Therefore, it is found that the spinning naked singularity may cause vacuum decay within the cosmological timescale with the change of singularity mass by $\alpha < 0.1 \%$ while the singularity would be covered by the event horizon due to the angular momentum transport.

\section{Conclusions}
Let us summarize the main results of this {\it Letter}. We first investigated a vacuum decay around a spinning naked singularity by using the Israel junction condition. Our calculations are valid when the nucleated bubble wall is thin compared to the radius of the bubble (thin wall approximation), and $r_{\text{s} \pm} \ll R \ll 1/H_{\pm}$ holds. Under these conditions, we found that the vacuum decay, by which all angular momentum of the seed naked singularity is transferred to the nucleated bubble wall, is most likely to occur. Such a vacuum decay involves the formation of event horizon around the singularity and the Bekenstein-Hawking entropy increases, which enhances the decay rate significantly. Fixing the parameters of our model motivated by the Higgs metastability \cite{Burda:2015yfa}, we also found that the small-mass naked singularities of $M_+ < 10^3 M_{\text{Pl}}$ may catalyze vacuum decays within the cosmological time scale and the singularity mass may change by $0.0001 \% \leq \alpha < 0.1 \%$, where the lower bound for $\alpha$ comes from the limitation of (\ref{condition1}). Although the small-mass naked singularities can be long-lived than the cosmic age in the astrophysical point of view \cite{1975AA4565D,deFelice:2007fs}, our scenario implies that the possible Higgs metastability itself prevents them to exist in the Universe (in the present Higgs vacuum state) in a quantum mechanical manner.

\textit{Acknowredgement.}---We thank Masaki Yamada and Masahide Yamaguchi for helpful comments and discussions. This work was supported by the Perimeter Institute for Theoretical Physics, and the JSPS Overseas Research Fellowships. Research at the Perimeter Institute is supported by the Government of Canada through Industry Canada, and by the Province of Ontario through the Ministry of Research and Innovation.

\appendix
{
\section{Error in the second Israel junction condition}
In this appendix, we estimate the error in the second Israel junction condition caused by a small discontinuity of the induced metric $\delta h_{\mu \nu}$.
The second Israel junction condition is
\begin{equation}\displaystyle
K_{\mu \nu}^{(+)} - K_{\mu \nu}^{(-)} = - 8 \pi G \lim_{\Delta \to 0} \int^{\Delta}_{- \Delta} (T_{\mu \nu} - \frac{1}{2} h_{\mu \nu} T) dx,
\label{app_1}
\end{equation}
where $h_{\mu \nu}$ is the induced metric, $T_{\mu \nu}$ is the energy momentum tensor in 4-dimension and $T$ is the trace of $T_{\mu \nu}$.
Let us suppose that there is a negligibly small discontinuity in $h_{\mu \nu}$. In this case, $h_{\mu \nu}$ can be expressed as
\begin{equation}
h_{\mu \nu} = h_{\mu \nu}^{(C)} + \delta H_{\mu \nu},
\end{equation}
where $h_{\mu \nu}^{(C)}$ and $\delta H_{\mu \nu}$ are the continuous and dis-continuous components and
\begin{equation}
\delta H_{\mu \nu} \sim \delta h_{\mu \nu} \Theta (x),
\end{equation}
where $\Theta (x)$ is the step function. Ginven $T = S \delta (x)$, the second term in the right hand side of (\ref{app_1}), can be written as
\begin{widetext}
\begin{align}\displaystyle
4 \pi G \lim_{\Delta \to 0} \int^{\Delta}_{-\Delta} h_{\mu \nu} T dx = 4 \pi G h_{\mu \nu}^{(C)} S + S \delta h_{\mu \nu} 4 \pi G \lim_{\Delta \to 0} \int^{\Delta}_{-\Delta} \Theta (x) \delta (x) dx = 4 \pi G h_{\mu \nu}^{(C)} S + 2 \pi G S \delta h_{\mu \nu}.
\end{align}
\end{widetext}
Therefore, a small discontinuity in the induce metric does not lead to singularity in the second Israel junction condition and it is suppressed when $\delta h_{\mu \nu}/ h_{\mu \nu}^{(C)} \ll 1$.
}

\end{document}